\documentclass[a4paper,10pt, onecolumn]{article}
\usepackage{lineno,hyperref}
\usepackage{graphicx,graphics}
\usepackage{amsmath}
\usepackage{amssymb}
\usepackage{epstopdf}
\usepackage{amstext}
\usepackage{amsthm}
\usepackage{amsfonts}
\usepackage{latexsym}
\usepackage{array}
\usepackage{xfrac}
\usepackage{color}
\usepackage{multirow}
\usepackage{fontenc}
\usepackage{bm}
\usepackage{adjustbox}
\usepackage{dcolumn}
\usepackage{color}
\usepackage{subcaption}
\usepackage{graphicx}
\usepackage{multicol}
\begin{document}

\title{\Large\bf Gogny forces in the astrophysical context}

\author{X. Vi\~nas, C. Gonzalez-Boquera, M. Centelles \\ \small Departament de F\'isica Qu\`antica i Astrof\'isica 
and Institut de Ci\`encies del Cosmos (ICCUB), \\ \small 
Facultat de F\'isica, Universitat de Barcelona, Mart\'i i Franqu\`es 1, E-08028 Barcelona, Spain
\and L.M. Robledo \\ \small Departamento de F\'isica Te\'orica,
Facultad de F\'isica, Universidad Aut\'onoma de Madrid,
\\ \small E-28049 Madrid, and Center for Computational Simulation,
Universidad Polit\'ecnica de Madrid, \\ \small 
Campus de Montegancedo, Boadilla del Monte, E-28660 Madrid, Spain,
\and C. Mondal \\ \small Departament de F\'isica Qu\`antica i Astrof\'isica 
and Institut de Ci\`encies del Cosmos (ICCUB), \\ \small 
Facultat de F\'isica, Universitat de Barcelona, Mart\'i i Franqu\`es 1, E-08028 Barcelona, Spain}
\date{}

\maketitle
\begin{abstract}
The most successful Gogny interactions of the D1 family, namely D1S, D1N and D1M,
suffer the common problem of a too soft neutron matter equation of state at high density, which 
prevents them from predicting a maximal mass of neutron stars of two solar masses, 
as required by recent astronomical observations. To cure this deficiency, we have proposed 
recently a reparametrization of the D1M force by fine tuning the slope of the 
symmetry energy in such a way that it preserves the ground-state properties of D1M in finite 
nuclei and also describes successfully the global properties of neutron stars, in particular its maximal 
mass, in consonance with the observational data. In this contribution we revisit this 
reparametrization by discussing two modified Gogny forces, dubbed D1M$^*$ and D1M$^{**}$.
     
\end{abstract}
\section{Introduction}
The Gogny forces 
were established by D. Gogny more than thirty years ago in order
to describe simultaneously the mean field and the pairing field 
with the same effective interaction \cite{decharge80}. These forces are specially well adapted to
study ground-state properties of spherical and deformed nuclei through the
Hartree-Fock-Bogoliubov (HFB) formalism in configuration space using harmonic 
oscillator (HO) basis. Large-scale calculations of this type \cite{CEA} have been
performed using the D1S parametrization of the Gogny force \cite{berger91},
which provide masses and radii as well as pairing and deformation properties
of finite nuclei in good agreement with the experimental data. However, it was found that there is
an energy drift in neutron-rich nuclei when computed with the D1S force \cite{pillet17}.
To improve this limitation, new parametrizations of the D1 family of interactions,
namely D1N \cite{chappert08} and D1M \cite{goriely09}, were proposed. It should
be mentioned that in order to calibrate these forces it is {required} that the 
D1N and D1M forces qualitatively reproduce the microscopic neutron matter equation 
of state (EOS) of Friedman and Pandharipande \cite{friedman81}. The parameters of the
D1N force were determined following the D1 fitting protocol \cite{decharge80}, while 
the parameters of the D1M interaction were adjusted to reproduce the experimentally
known masses of 2149 nuclei. This fit of the D1M force, which also takes into account the 
quadrupole correlation energies, predicts the experimental masses of the previously
mentioned nuclei with a {\it rms} deviation of only 798 keV \cite{goriely09}.

In spite of this good description of ground-state properties of finite nuclei,
the extrapolation to the neutron star domain is not completely satisfactory. For example, as
discussed in detail in Refs. \cite{loan11,sellaheva14,gonzalez17,gonzalez18},
the successful Gogny forces of the D1 family, which nicely reproduce the ground-state
properties of finite nuclei, namely D1S, D1N and D1M, are unable to reach a maximal neutron
star (NS) mass of 2$M_{\odot}$ as required by recent astronomical observations 
\cite{demorest10,antoniadis13} and only the D1M interaction predicts a NS mass 
above the canonical value of 1.4$M_{\odot}$ \cite{sellaheva14,gonzalez17,gonzalez18}. 
This can be observed in the panel a) of Figure 1 where the mass-radius relations 
computed with different mean field models, including the most successful Gogny interactions, 
are displayed. 

Some unique properties of the Gogny forces, such as its finite range and the fact that
the particle-particle and the particle-hole interactions can be treated with the same force,
may also have an important role in the astrophysical context. To achieve this goal, we have built two
new Gogny interactions, which we call D1M$^*$ and D1M$^{**}$, performing a modification of the
parameters of D1M in such a way that these forces reproduce finite nuclei data 
 in as good agreement as those obtained using D1M and, at the same time, provide a description of NS at the same
level of some Skyrme forces, such as SLy4 \cite{chabanat98}, specially designed for the study of NS.      
  
The paper is organized as follows. The first section is devoted to the fitting procedure of 
two new interactions namely D1M$^*$ and D1M$^{**}$ and to the discussion of some relevant predictions of these forces in the context 
of neutron stars. In the second section the ability of the D1M$^*$ interaction is analyzed by describing some selected 
properties of finite nuclei, comparing them with the predictions from the D1M force. 
Finally our conclusions are laid in the last section.

\section{The D1M$^*$ and D1M$^{**}$ interactions}
The standard Gogny interaction of the D1 family consists of a finite range part, which is modeled by two
Gaussian terms that include all the possible spin-isospin exchange terms, plus a zero-range
density-dependent term. Including the spin-orbit force, which is also of contact type, the Gogny force
\cite{decharge80} reads:
\begin{eqnarray}\label{VGogny}
  V (\mathbf{r}_1 , \mathbf{r}_2) &=& \sum_{i=1,2}
  \big( W_i + B_i P_{\sigma} - H_i P_{\tau} - M_i P_{\sigma}P_{\tau}\big)
   e^{-\frac{r^2}{\mu_i^{2}}} \nonumber
   \\
  && + t_3 (1+ x_3 P^\sigma) \rho^\alpha (\mathbf{R}) \delta (\mathbf{r}) \nonumber
 \\
  && + i W_{LS} (\sigma_1 + \sigma_2) (\mathbf{k'} \times \delta (\mathbf{r}) \mathbf{k}),
\end{eqnarray}
where ${\bf r}={\bf r_1}-{\bf r_2}$ and ${\bf R}=({\bf r_1}+{\bf r_2})/2$ are the relative and the center 
of mass coordinates, and $\mu_1 \simeq$0.5-0.7 fm and $\mu_2 \simeq$1.2 fm are the ranges of the two 
Gaussian form factors, which simulate the short- and long-range components of the force, respectively.

\begin{table}[b!]
\centering
\begin{tabular}{lccccccc}
\hline
         &$\rho_0$ & $E_0$ & $K$ & $m^*/m$   & $E_{\rm sym}(\rho_0)$ & $E_{\rm sym}(0.1)$ & $L$  \\
         & (fm$^{-3}$) &  (MeV)  &  (MeV) &        & (MeV) &(MeV) & (MeV)\\  \hline
D1M$^*$  &       0.1650          & $-$16.06  & 225.4   &  0.746  & 30.25   & 23.82 & 43.18  \\
D1M$^{**}$  &    0.1647          & $-$16.02  & 225.0   &  0.746  &29.37   & 23.80 & 33.91  \\
D1M      &       0.1647          & $-$16.02  & 225.0   &  0.746  & 28.55   & 23.80 & 24.83  \\
D1N      &       0.1612          & $-$15.96  & 225.7   &  0.697  & 29.60   & 23.80 & 33.58 \\
D1S      &       0.1633          & $-$16.01  & 202.9   &  0.747  & 31.13   & 25.93 & 22.43 \\
D2       &       0.1628          & $-$16.00  & 209.3   &  0.738  & 31.13   & 24.32 & 44.85  \\
SLy4     &       0.1596          & $-$15.98  & 229.9   &  0.695  & 32.00   & 25.15 & 45.96 \\\hline
\end{tabular}
\caption{Nuclear matter properties predicted by the D1M$^{*}$, D1M$^{**}$, D1M, D1N, D1S and D2 Gogny
interactions and the SLy4 Skyrme force.}
\label{inm}
\end{table}

The standard nuclear matter properties predicted by some relevant Gogny forces are displayed 
in Table 1. From this Table we can see that the different properties of symmetric nuclear matter (SNM) {\it e.g.}
the saturation density $\rho_0$, the binding energy per nucleon at
saturation $E_0$, the incompressibility $K$ and the effective mass $m^*/m$ predicted by the 
different forces considered are rather similar. This is because the properties of SNM  
at low densities are well constrained by the properties of terrestrial 
nuclei. However, the symmetry energy and, in particular, its slope at saturation, which governs the 
isovector sector, differs more among the different models. We can see that some forces (the new D1M$^*$,
D1M$^{**}$ and D2 Gogny interactions and the SLy4 Skyrme force), which predict maximum mass of NS around 2$M_{\odot}$ (see Figure 1), have 
a slope of the symmetry energy $L$ about 45 MeV, while other forces, such as D1S, D1N and D1M, 
which have smaller values of the slope parameter $L$, are unable to reach a maximum mass of NS in excess of
2$M_{\odot}$ as it can be seen from panel a) of Figure 1. From panel b) of Figure 1 we can also observe 
that those forces which are able to reach a maximal mass of  2$M_{\odot}$ predict 
a central density around seven times the
saturation density while forces predicting smaller maximal mass have larger central densities.   
 
The symmetry energy is defined as $E_\mathrm{sym}(\rho)=\frac{1}{2}\partial^2 E_b(\rho,\delta)/\partial \delta^2 \vert_{\delta=0}$,
where $E_b(\rho,\delta)$ is the energy per particle in asymmetric nuclear matter of density $\rho=\rho_n+\rho_p$
and asymmetry $\delta=(\rho_n-\rho_p)/\rho$, $\rho_n$ and $\rho_p$ being the neutron and proton densities,
respectively. The symmetry energy can be understood as the energy cost for converting all protons in neutrons in 
symmetric nuclear matter. Therefore its slope at the saturation density practically corresponds to the slope of the 
neutron EOS at this density and, consequently, has an important impact on the behaviour of the NS EOS
at densities above saturation. The slope parameter is defined as 
$L=3\rho_0 \frac{\partial E_\mathrm{sym}(\rho)}{\partial \rho}\vert_{\rho=\rho_0}$, where $\rho_0$ is the saturation density.
This parameter $L$  is also connected to different properties of finite nuclei, as for example the neutron 
skin thickness in a heavy nucleus such as $^{208}$Pb (see \cite{centelles09,vinas14} and references therein). 
The symmetry energy as a function of the density is displayed for several Gogny interactions in panel a) of
Figure 2. At subsaturation densities the symmetry energy computed with all the forces show a similar
behaviour and takes a value around 30 MeV at saturation (see Table 1). In this regime the symmetry 
energy falls within, or lies very close, to the region constrained by the Isobaric Analog States 
\cite{danielewicz14}, which implies that the symmetry energy predicted by these Gogny interactions is 
well constrained by finite nuclei data. Above saturation, the behaviour of the symmetry energy is 
strongly model dependent. From this Figure two different patterns can be observed. On the one hand,
the symmetry energy computed with the  D1S, D1N and D1M interactions increases till reaching a maximum
value around 30-40 MeV and then bends and decreases with increasing density until vanishing at some
density where the isospin instability starts. On the other hand, the other forces, namely D1M$^*$, D1M$^{**}$
and D2, predict a symmetry energy with a well defined increasing trend with growing density. This different
behaviour of the symmetry energy strongly influences the EOS of the NS matter (which for simplicity we assume 
made of neutrons, protons and electrons in charge and $\beta$-equilibrium) as it can be seen in the panel b)  of
Figure 2. From this panel we observe that except the EOS obtained with the D1S interaction, the EOSs predicted 
by the other forces grow with increasing density. However, not all forces with an increasing EOS in NS 
matter are able to predict a NS with maximum mass of 2$M_{\odot}$ or above because these EOSs are too
soft at high densities several times the saturation density.

\begin{figure}[t]
\centering
\includegraphics[clip=true,scale=0.3,width=0.50\columnwidth]{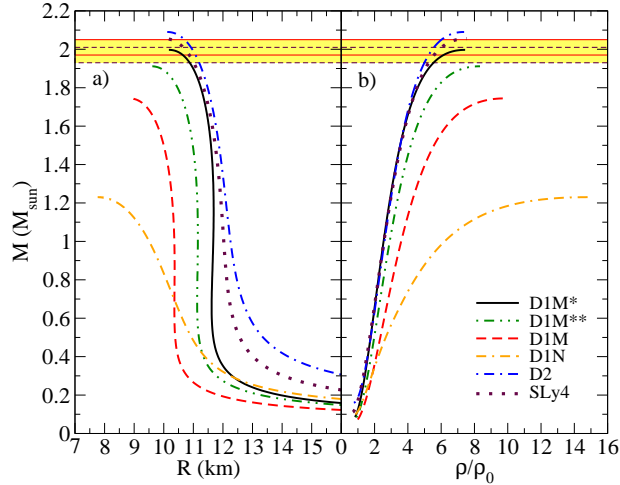}
\caption{a) Mass-Radius relation for different Gogny forces compared with the prediction of the 
SLy4 Skyrme force, which is taken as a benchmark. The constraint on the observed maximum mass of neutron 
stars \cite{demorest10,antoniadis13} is depicted as the horizontal shaded region.
b) Mass of neutron stars predicted by different Gogny forces and the SLy4 interaction as a function 
of the central density (in units of the saturation density).}\label{f01}
\end{figure}
In Figures 1 and 2, in addition to the standard Gogny forces D1S, D1N and D1M, we have also considered the 
D1M$^*$, D1M$^{**}$ and D2 interactions. The D1M$^*$ force is a reparametrization of the D1M interaction 
introduced in Ref. \cite{gonzalez18}, which leaves the description of finite nuclei with an average level of quality similar 
to that provided by D1M and, at the same time, is also able to predict a maximum mass of the NS of 2 $M_{\odot}$. The D1M$^{**}$ is another 
Gogny force built up as D1M$^*$ but predicting a maximum mass of NS {about} 1.9$M_{\odot}$, which corresponds to the 
lower value, within the error bars, of the heaviest observed masses \cite{demorest10,antoniadis13}. 
The D2 interaction is a new Gogny force, devised by the Bruy\`eres-le-Ch\^atel group, which instead of the standard 
zero-range density-dependent term of the D1 family of Gogny forces, contains a density-dependent finite-range term 
\cite{chappert07,chappert15}. This force is fitted according to the D1 protocol \cite{decharge80} including a qualitative 
reproduction of the Friedman and Pandharipande EOS. This new force is free of the energy drift for exotic nuclei observed 
in D1S, but, as it is pointed out in \cite{chappert15}, the description of the nuclear masses has not reached yet the 
quality obtained with the D1M force.     
\begin{figure}[t]
\centering
\includegraphics[clip=true,scale=0.3,width=0.50\columnwidth]{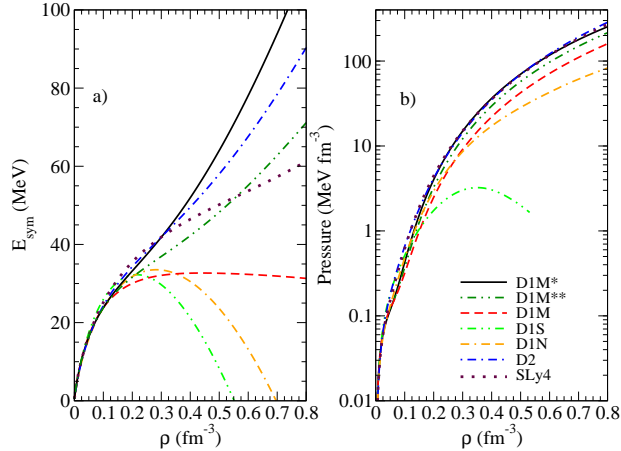}
\caption[]{a) Symmetry energy computed with the D1M$^{*}$, D1M$^{**}$, D1M, D1N, D1S and D2
Gogny interactions and the SLy4 Skyrme force as a function of the density.
b) EOS of $\beta$-stable neutron star matter as a function of the density predicted by the same forces as in  
panel a).}\label{f02}
\end{figure}

\subsection{The fitting procedure}

\begin{table}[b!]
\centering
\begin{tabular}{c|rrrrr}
\hline
D1M      & \multicolumn{1}{c}{$W_i$} & \multicolumn{1}{c}{$B_i$} & \multicolumn{1}{c}{$H_i$} & \multicolumn{1}{c}{$M_i$} & \multicolumn{1}{c}{$\mu_i$}  \\ \hline
$i$=1    & -12797.57  & 14048.85   & -15144.43  & 11963.81   & 0.50 \\
$i$=2    &    490.95  &  -752.27   &    675.12  &  -693.57   & 1.00  \\\hline
D1M$^*$     & \multicolumn{1}{c}{$W_i$} & \multicolumn{1}{c}{$B_i$} & \multicolumn{1}{c}{$H_i$} & \multicolumn{1}{c}{$M_i$} & \multicolumn{1}{c}{$\mu_i$}  \\ \hline
$i$=1    & -17242.0144 & 19604.4056  & -20699.9856 & 16408.3344 & 0.50 \\
$i$=2    &    712.2732 &  -982.8150  &    905.6650 &  -878.0060 & 1.00 \\\hline
D1M$^{**}$     & \multicolumn{1}{c}{$W_i$} & \multicolumn{1}{c}{$B_i$} & \multicolumn{1}{c}{$H_i$} & \multicolumn{1}{c}{$M_i$} & \multicolumn{1}{c}{$\mu_i$}  \\ \hline
$i$=1    & -15019.7922 & 16826.6278 & -17922.2078 & 14186.1122 & 0.50 \\
$i$=2    &    583.1680 &  -867.5425  &   790.3925 &  -785.7880 & 1.00 \\\hline
\end{tabular}
\caption{Parameters of the D1M, D1M$^*$ and D1M$^{**}$ Gogny interactions, where $W_i$, $B_i$, $H_i$ and
$M_i$ are in MeV and $\mu_i$ in fm. The coefficients $x_3=1$, $\alpha=1/3$ and $W_{LS}=115.36$ MeV fm$^5$ 
are the same in the three interactions, and $t_3$ has values of $t_3=1562.22$ MeV fm$^4$ for the Gogny D1M and D1M$^{**}$
 forces and $t_3=1561.22$ MeV fm$^4$ for the D1M$^*$ interaction.}
\label{param}
\end{table}
The new Gogny interaction D1M$^*$ is obtained starting from the D1M force and performing a
controlled change of its parameters. This means that we only modify the finite-range spin-isospin
strength coefficients keeping the ranges of the two Gaussian form factors as well as the
zero-range density dependent part of the force with the same values as in the original
D1M force.
In order to determine the finite-range parameters of the new Gogny interaction D1M$^*$, we proceed in a
similar way to {those} used in previous literatures to generate families of Skyrme interactions or RMF parametrizations,
 {\it viz.} SAMi-J\cite{roca13}, KED0-J \cite{agrawal05} or TAMU-FSU \cite{piekarewicz11,fattoyev13}.
The basic idea to obtain these families is the following. Starting from a well calibrated and successful mean-field
model, one modifies the values of some parameters, which determine the symmetry energy, around their optimal values
retaining as much as possible values of the binding energies and radii of finite nuclei of the original model.
Four of the eight initially free parameters of the force, namely $W_i$, $B_i$, $H_i$ and $M_i$ ($i$=1,2), are constrained
by imposing that the saturation density, energy per particle, incompressibility modulus and effective mass in symmetric
nuclear mater take the values corresponding to the original D1M interaction. In order to have a right behaviour of the
asymmetric nuclei, we also impose that the symmetry energy in uniform matter calculated at a density of 0.1 fm$^{-3}$
computed with D1M and D1M$^*$ to be the same. The reason of this constraint is based on the empirical law of
Ref. \cite{centelles09}, which demonstrates that the symmetry energy at some subsaturation density about 0.1 fm$^{-3}$,
%for a heavy nucleus such as $^{208}$Pb, 
calculated with a given nuclear force, coincides with the symmetry energy of $^{208}$Pb nucleus calculated with the Droplet
Model \cite{myers69} for the same force, which contains both, bulk and surface contributions. To preserve the pairing properties of D1M
in the S=0 T=1 channel, we also impose that in the new parametrization D1M$^*$ the combination of parameters $W_i-B_i-H_i+M_i$
($i$=1,2) take the same values to those in the original force D1M.
With this protocol, we are able to determine seven of the eight initially free parameters of D1M$^*$ as a function of the
eighth parameter, which we chose to be $B_1$. This free parameter is used to modify the slope of the symmetry energy at
saturation and therefore, the behaviour of the neutron matter EOS above saturation, which in turn determines the maximum mass
of the NS by solving the Tolman-Oppenheimer-Volkoff (TOV) equations. In this way the parameters corresponding to the finite 
range part of the new D1M$^*$ interaction are completely determined. We observe, however, that the description of nuclear 
masses provided by the D1M$^*$ force  degrades slightly as compared with predictions of the original D1M interaction. 
To correct this deficiency, we perform in D1M$^*$ an additional small refit of the zero-range strength $t_3$ of about 
1 MeV fm$^4$ to recover the same mass {\it rms} value to that using the D1M parametrization. The fitting protocol of the 
D1M$^{**}$ is the same only that the required maximum mass of a NS is now 1.9 $M_{\odot}$ instead of 2 $M_{\odot}$ as in 
the case of D1M$^*$ (see below).
 
The parameters of the new forces D1M$^*$ and D1M$^{**}$ are collected in Table 2 together with the ones of the
original D1M interaction. We observe that the change of the finite range parameters, as compared with the original
ones of the D1M force, is larger for the D1M$^*$ force than for the D1M$^{**}$ interaction because the variation in the
isovector sector is more important in the former than in the latter force. Looking at the D1M family in Table 1, we see
that the only nuclear matter property that shows a relevant modification is the slope of the symmetry energy, which changes
from 24.83 MeV (D1M) to 33.91 MeV (D1M$^{**}$) and 43.18 MeV (D1M$^*$). The symmetry energy
at saturation also changes but in a much less extent. All the remaining nuclear matter properties of the forces of the D1M family displayed in Table 1 take the
same values as a consequence of our fitting procedure. Looking at Figure 1 and keeping in mind the values of the slope
of the symmetry energy of the D1M family, it can be seen that when the slope parameter of the force increases, the maximum
mass of the NS predicted by the force also increases. This is, however, only a qualitative rule and exceptions may appear.
For example D1N and D1M$^{**}$ have almost the same slope of the symmetry energy at saturation but the maximum mass predicted
by the former is around 1.2 $M_{\odot}$ and 1.9 $M_{\odot}$ for the latter. This fact points out that, in spite of the same
slope of the symmetry energy at saturation, the behaviour of the symmetry energy above saturation (see upper panel of Figure 2)
strongly determines the EOS at high density and therefore the maximum NS mass predicted by each force.

\section{Finite nuclei properties described with the D1M$^*$ force}

One of the goals of parametrizing D1M$^*$ and D1M$^{**}$ is to reproduce nuclear structure 
properties of finite nuclei with the same global quality as the original D1M force. The finite nuclei
calculations have been performed with the computer code HFBaxial \cite{robledo02} which solves the HFB
equations in a HO basis using the Gogny interaction. Large-scale HFB calculations using this code and
the D1M Gogny force have been done some time ago \cite{robledo11}. More details about the technicalities 
for solving the HFB equations and the minimization procedure for finding the ground-state properties have been 
reported in \cite{gonzalez18}. Although the calculations of finite nuclei properties with the D1M$^{**}$ 
force have not been performed as extensively as with D1M$^{*}$, looking at their parameters reported in Table \ref{param}, it is expected that 
the predictions of D1M$^{**}$ will lie between the ones of D1M and D1M$^{*}$. Our preliminary investigation confirms this expectation.
\begin{figure}[t]
\centering
\includegraphics[clip=true,width=0.46\columnwidth]{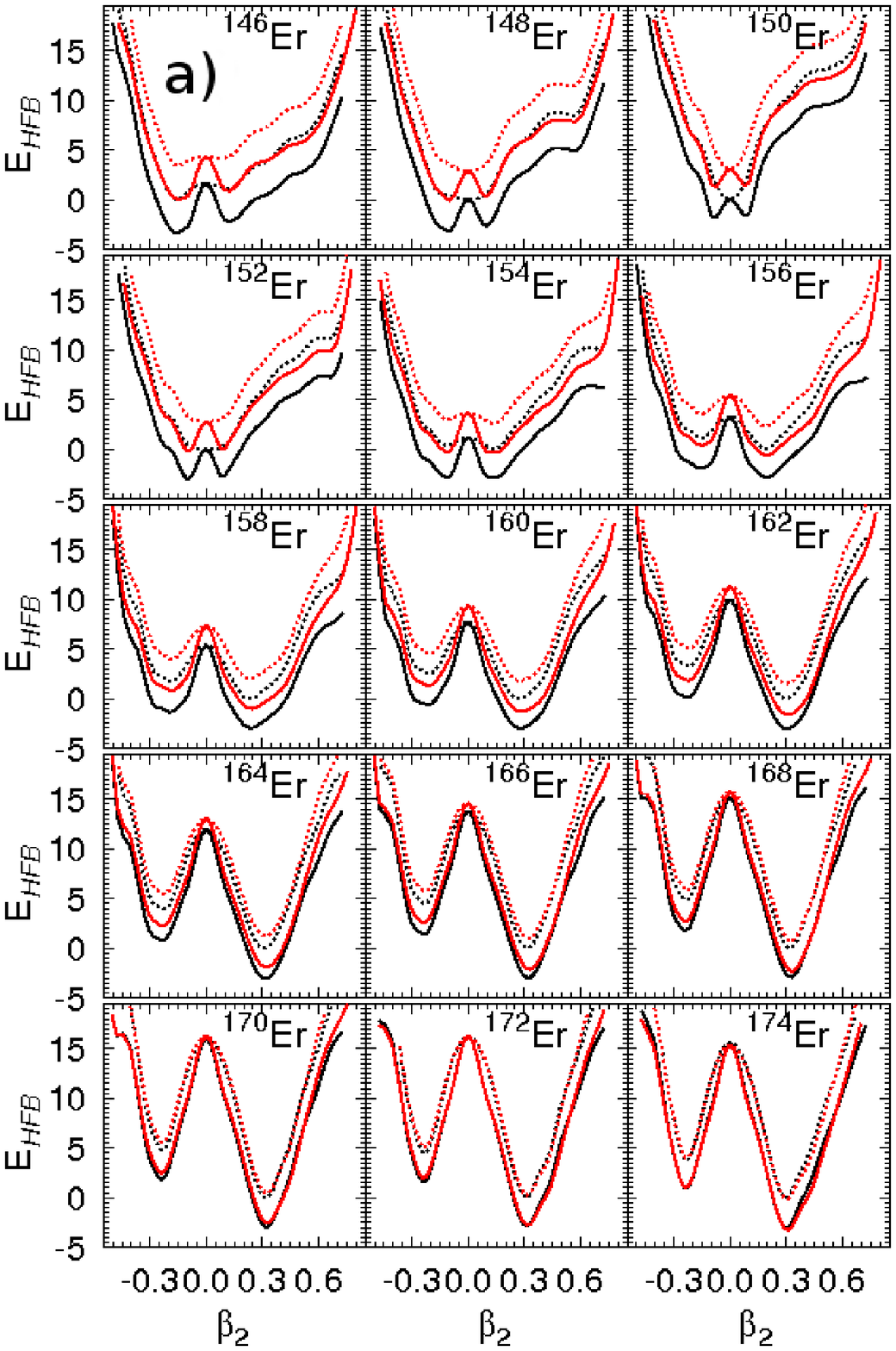}
\includegraphics[clip=true,width=0.50\columnwidth]{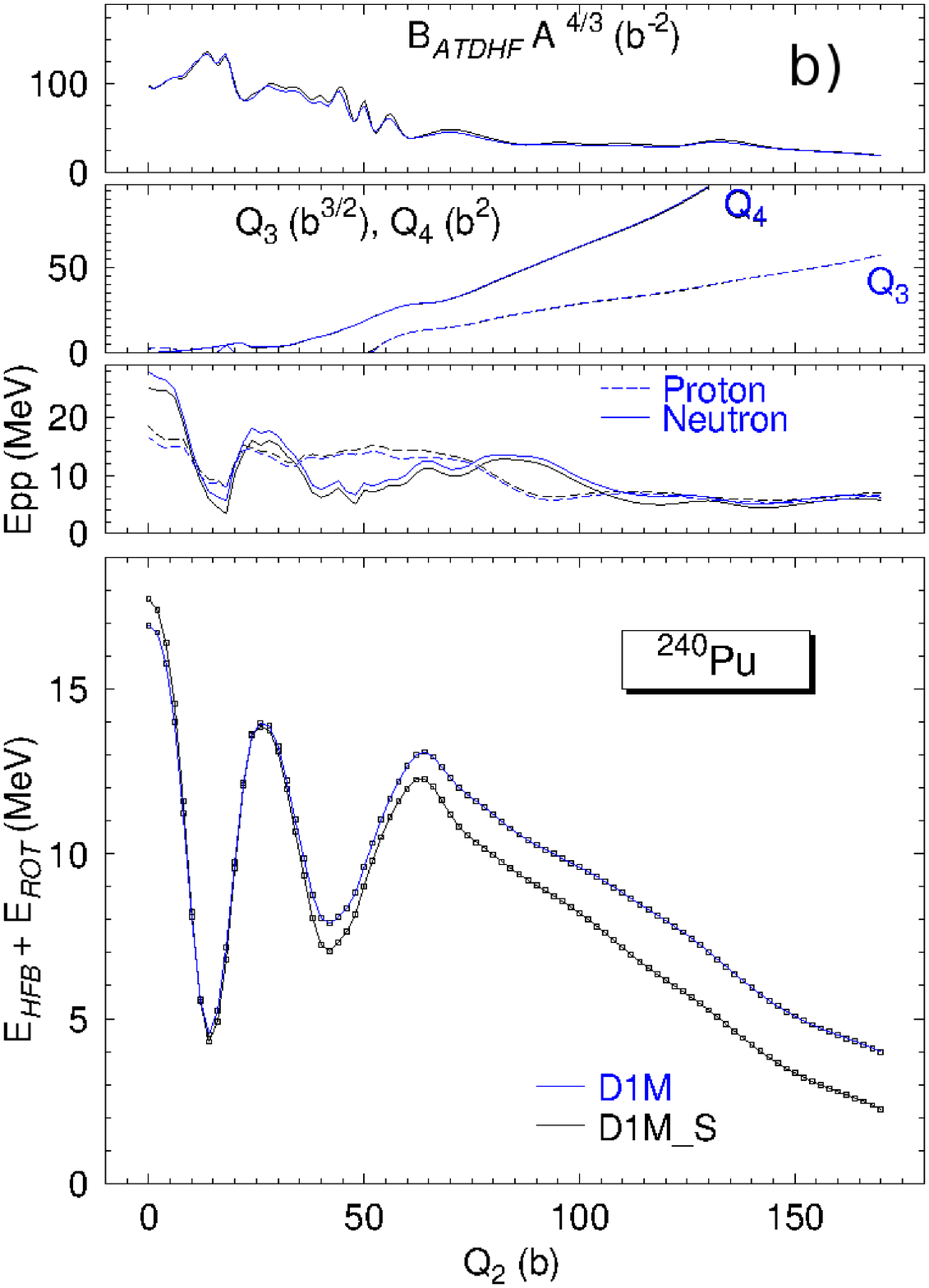}
\caption[]{a) Potential energy surfaces of the Er isotopic chain computed with the D1M (red) and 
D1M$^*$ (black) interactions as a function of the quadrupole deformation parameter $\beta_2$. Solid and dotted 
curves include and do not include the rotational energy. b) Fission barrier of the nucleus $^{240}$Pu 
as a function of the quadrupole moment $Q_2$ calculated with the same Gogny forces. The evolution of the mass 
parameter, octupole and hexadecapole moments and neutron and proton pairing energies along the fission path 
are also displayed in the same Figure.}\label{f03}
\end{figure}
\subsection{Binding energies and neutron and proton radii}

Fine tuning of the $t_{3}$ parameter of D1M$^{*}$ has been performed by
minimizing the {\it rms} deviation, $\sigma_{E}$, of the binding energies 
of 620 even-even nuclei \cite{audi12}. The obtained value of 1.34 MeV is
very close to the 1.36 MeV obtained for D1M under the same circumstances.  
This indicates a similar performance of both parametrizations in the average 
description of binding energies along the periodic table.
The differences between the theoretical binding energies calculated with the D1M$^*$
force and the experimental values are scattered around zero and do not show any drift with increasing 
neutron number. The agreement between theory and experiment is good for medium and heavy nuclei and
deteriorates for light nuclei, as usually happens with mean field models of different nature. This can 
be seen  in the upper panel of Figure 5 of Ref. \cite{gonzalez18}. The differences between the binding
energies computed with the D1M$^*$ and D1M are never larger than $\pm 2.5$ MeV and show a clear shift along
isotopic chains because of the different density dependence of the symmetry energy in both forces,
as displayed in the lower panel of Figure 5 of Ref. \cite{gonzalez18}.
The slope parameter $L$ predicted by the D1M$^*$ force is larger than the one of D1M, as can be 
seen in Table 2. As a consequence \cite{centelles09}, it is expected that for neutron-rich nuclei 
the radius of the neutron distribution calculated with D1M$^*$ is larger than the value obtained 
with D1M. On the other hand, and due to the fitting protocol to obtain the different members of the
D1M family, it is also expected that the  radii for the proton distribution calculated with the D1M and D1M$^*$ 
forces are roughly the same. The HFB calculations performed with these interactions confirm this expectation.

\subsection{Potential Energy Surfaces}
An important aspect of any nuclear interaction is the way  it determines the response of the
nucleus to shape deformation, in particular to the quadrupole deformation. To know if a nucleus is quadrupole 
deformed or not, plays a crucial role in the determination of the low energy spectrum. To study the 
response of the D1M$^*$ force to the quadrupole deformation, we have performed constrained HFB calculations 
in finite nuclei fixing the quadrupole moment $Q_{20}$ to given values, which allow to obtain the potential energy
surfaces (PES). As an example, in panel a) of Figure 3 the PES along the Er isotopic chain is displayed as a function of the deformation parameter 
$\beta_2$ for the original D1M interaction and for the modified D1M$^*$ force.
From this Figure, it can be observed that the curves corresponding to the calculations performed with the D1M 
and D1M$^*$ forces follow basically the same trends  with a small parallel displacement of one curve with respect to the 
other.
  
\subsection{Fission Barriers}
Finally we discuss the fission barrier of the paradigmatic nucleus $^{240}$Pu, which is displayed in the
right bottom panel of Figure 3. We see that the inner fission barrier predicted by D1M and D1M$^*$ is the same 
in both models with a value $B_I$=9.5 MeV. This value is a little bit large compared with the ``experimental" 
value of 6.05 MeV. However, it should be pointed out that triaxiality effects, not accounted for in the present 
calculation, might lower the inner barrier by 2-3 MeV. The excitation energy of the fission isomer $E_{II}$ is
3.36 MeV computed with D1M and 2.80 MeV with D1M$^*$. The outer fission barrier $B_{II}$ height are 
8.58 and 8.00 MeV calculated with the D1M and D1M$^*$ forces, respectively. These values clearly 
overestimate the empirical value, which is 5.15 MeV. In the other panels of the same Figure we have also 
displayed as a function of the quadrupole deformation the neutron and proton pairing energies, the octupole 
moment (responsible for asymmetric fission) and the hexadecapole moment of the mass distributions as well as
the collective inertia. All these quantities take very similar values computed with both interactions.

\section{Conclusions}
In this paper we have revisited some reparametrizations of the  Gogny D1M force, which preserves
most of its relevant properties in finite nuclei, and is still able to predict a stiffer neutron 
equation of state by a suitable modification of the slope of the symmetry energy. 
This enables one to increase the 
predicted maximal mass of neutron stars. We propose two different reparametrizations, namely D1M$^*$ and
D1M$^{**}$, which predict maximal mass of neutron stars of 2 and 1.9 solar masses, respectively,  in agreement 
with the range of values provided by recent astronomical observations. With these modified interactions we study 
some basic properties of finite nuclei, such as binding energies, neutron and proton radii, response to quadrupole 
deformation and fission barriers.  We find that both, D1M$^*$ and D1M$^{**}$, perform as well as D1M in all the 
concerned properties of finite nuclei. We have also verified that the description of 
neutron star properties by these new forces are very similar to those obtained with the Skyrme SLy4 interaction, which is 
designed specially for working in the astrophysical scenario. To summarize, one can conclude
that the D1M$^*$ and D1M$^{**}$ forces are a good alternative to describe simultaneously finite nuclei and neutron stars providing results
in harmony with the experimental and observational data.

\section{Acknowledgments}
The work of L.M.R. was supported by Spanish Ministry of Economy and Competitiveness (MINECO)
Grants No FPA2015-65929-P and FIS2015-63770-P.
C.G., M.C., X.V. and C.M. were partially supported by Grant FIS2017-87534-P from MINECO and FEDER, 
and Project MDM-2014-0369 of ICCUB (Unidad de Excelencia Mar\'{\i}a de Maeztu) from MINECO.
C.G. also acknowledges Grant BES-2015-074210 from MINECO.


\begin{thebibliography}{99}

\bibitem{decharge80} J. Decharg\'e and D. Gogny, \textit{Phys.Rev.C}
\textbf{21} (1996) 1568.

\bibitem{CEA} CEA web page {\it www-phynu.cea.fr}.

\bibitem{berger91} J.F. Berger, M. Girod and D. Gogny, 
\textit{Comp.Phys.Commun.} \textbf{63} (1991) 305.

\bibitem{pillet17}
N. Pillet and S. Hilaire, \textit{The European Physical Journal A} \textbf{53} (2017) 193.

\bibitem{chappert08} F. Chappert, M. Girod and  S. Hilaire,
\textit{Phys.Lett.B} \textbf{688} (2008) 420.

\bibitem{goriely09} S. Goriely, S. Hilaire, M. Girod and S. P\'eru,
\textit{Phys.Rev.Lett.} \textbf{102} (2009) 242501.

\bibitem{friedman81} B. Friedman and V. Pandharipande,
\textit{Nuc.Phys.A} \textbf{361} (1981) 502.

\bibitem{loan11} D.T. Loan, H.H. Tan, D.T. Khoa and J. Margueron,
\textit{Phys.Rev.C} \textbf{83} (2011) 065809.

\bibitem{sellaheva14} R. Sellahewa and A. Rios,
\textit{Phys.Rev.C} \textbf{90} (2014) 054327.

\bibitem{gonzalez17} C. Gonzalez-Boquera, M. Centelles, X. Vi\~nas
and A. Rios, \textit{Phys.Rev.C} \textbf{96} (2017) 065806

\bibitem{gonzalez18} C. Gonzalez-Boquera, M. Centelles, X. Vi\~nas
and L.M. Robledo, \textit{Phys.Lett.B} \textbf{779} (2018) 195.

\bibitem{demorest10} P.B. Demorest, T. Pennucci, S.N. Ransom, 
M.S.E. Roberts and J.W.T. Hessels, \textit{Nature} \textbf{467} (2010) 1081.

\bibitem{antoniadis13} J. Antoniadis et al.,
\textit{Science} \textbf{340} (2013) 6131.

\bibitem{chabanat98} E. Chabanat, P. Bonche, P. Haensel, J. Meyer and 
R. Schaeffer, \textit{Nuc.Phys.A} \textbf{635} (1998) 441.

\bibitem{centelles09} M. Centelles, X. Roca-Maza, X. Vi\~nas and M. Warda,
\textit{Phys.Rev.Lett.} \textbf{102} (2009) 122502.

\bibitem{vinas14} X. Vi{\~{n}}as, M. Centelles, X. Roca-Maza and M. Warda, 
\textit{The European Physical Journal A} \textbf{50} (2014) 27.

\bibitem{myers69} W.D. Myers and W. Swiatecki,
\textit{Annals of Physics} \textbf{55} (1969) 395.

\bibitem{danielewicz14} P. Danielewicz and J. Lee,
\textit{Nuc.Phys.A} \textbf{922} (2014) 1.

\bibitem{chappert07} F. Chappert, Ph.D. thesis, Niversit\'e Paris-Sud XI, 2007,
https://tel.archives-ouvertes.fr/tel-00177379/en/.

\bibitem{chappert15} F. Chappert, N. Pillet, M. Girod, J.-F. Berger,
\textit{Phys.Rev.C} \textbf{91} (2015) 034312.

\bibitem{roca13}
X. Roca-Maza et al,
\textit{Phys.Rev.C} \textbf{87} (2013) 034301.

\bibitem{agrawal05}
B.K. Agrawal, S. Shlomo and V.K. Au,
\textit{Phys.Rev.C} \textbf{72} (2005) 014310.

\bibitem{piekarewicz11}
J. Piekarewicz,
\textit{Phys.Rev.C} \textbf{83} (2011) 034319.

\bibitem{fattoyev13}
F.J. Fattoyev and J. Piekarewicz,
\textit{Phys.Rev.Lett.} \textbf{111} (2013) 162501.

\bibitem{robledo02}
L.M. Robledo, HFBaxial computer code 2002.

\bibitem{robledo11}
L.M. Robledo and G.F. Bertsch,
\textit{Phys.Rev.C} \textbf{84} (2011) 054302. 

\bibitem{audi12} 
G. Audi, M. Wangh, A. Waspra, F. Kondev, M. MacCormick, X. Xu
and B. Pfeiffer, \textit{Chin.Phys.C} \textbf{36} (2012) 1287.

\end{thebibliography}
\end{document}